\documentclass[aps,prd,eprint,numerical,showpacs,showkeys,10pt]{revtex4-2}

\usepackage{longtable}
\usepackage{amsmath}
\usepackage{amssymb}
\usepackage{amsfonts}
\usepackage{graphicx}
\usepackage{dcolumn}
\usepackage{bm}
\usepackage{hyperref}
\hypersetup{colorlinks=false}

\begin{document}

\title[Quasibound states in the consistent 4DEGB gravity]{Quasibound States, Stability and Wave Functions of the Test Fields in the Consistent 4D Einstein--Gauss--Bonnet Gravity}

\date{\today}

\author{H. S. Vieira}
\email{horacio.santana.vieira@hotmail.com}
\email{horacio_vieira@ufla.br}
\email{horacio.santana-vieira@tat.uni-tuebingen.de}
\email{horacio.santana.vieira@ifsc.usp.br}
\affiliation{Department of Physics, Institute of Natural Sciences, Federal University of Lavras, 37200-000 Lavras, Brazil}
\affiliation{Theoretical Astrophysics, Institute for Astronomy and Astrophysics, University of T\"{u}bingen, 72076 T\"{u}bingen, Germany}
\affiliation{S\~{a}o Carlos Institute of Physics, University of S\~{a}o Paulo, 13560-970 S\~{a}o Carlos, Brazil}

\begin{abstract}
We examine the interaction between quantum test particles and the gravitational field generated by a black hole solution that was recently obtained in the consistent 4-dimensional Einstein--Gauss--Bonnet gravity. While quasinormal modes of scalar, electromagnetic, and Dirac fields have been recently studied in this theory, there is no such study for the quasibound states. Here, we calculate the spectrum of quasibound states for the test fields in a spherically symmetric and asymptotically flat black hole solution in the consistent 4-dimensional Einstein--Gauss--Bonnet gravity. The quasispectrum of resonant frequencies is obtained by using the polynomial condition associated to the general Heun functions. We also discuss the stability of the systems for some values of the Gauss-Bonnet coupling constant.
\end{abstract}

\pacs{02.30.Gp, 03.65.Ge, 04.20.Jb, 04.62.+v, 04.70.-s, 04.80.Cc, 47.35.Rs, 47.90.+a}

\keywords{alternative theories of gravity; covariant equations of motion; general Heun function; quasistationary level; wave eigenfunction}

\preprint{Universe.9.205(2023)}

\maketitle


%
%
\section{Introduction}\label{Introduction}
The quasibound states, quasinormal modes, and shadows of black holes are among the most interesting characteristic of such an astrophysical objects in the observational (measurable) spectra. In addition, some physical phenomena have been observed in modern experiments involving condensate matter and optical systems. In order to interpret these data, it is used the general theory of relativity. However, since some fundamental questions, as for example, the issues related to the quantum gravity phenomenon \cite{PhysRevLett.55.2656,PhysRevLett.116.061102,PhysLettB.756.350,GenRelGrav.50.49,AstrophysJ.875.L1}, cannot be solved with the Einstein's theory, some alternative theories of gravity have been proposed. Among these alternative approaches, we can mention the f(R), the Lovelock, and the Einstein--Gauss--Bonnet theories of gravity, where the last two deals with higher curvature corrections \cite{PhysRevD.98.024042,IntJModPhysD.26.1730001}.

The Einstein--Gauss--Bonnet (EGB) theory is one of the most promising approaches developed to deal with the higher curvature corrections that appears in the standard Einstein's theory (see Ref.~\cite{PhysRevLett.125.149001} and references therein). The EGB theory is quadratic in the curvature and leads to non-trivial corrections of the equation of motion when the Gauss-Bonnet (GB) term is coupled to a matter field, which can be, for instance, a dilaton. On the other hand, Aoki--Gorji--Mukohyama (AGM) \cite{PhysLettB.810.135843} developed the so-called consistent theory of 4-dimensional (4D) EGB gravity, where they used the Arnowitt--Deser--Misner (ADM) decomposition \cite{PhysRev.116.1322} to construct the Hamiltonian and then this theory has not infinite coupling.

In the present paper, we will use a black hole metric that is an exact solution of the (well-defined truly) 4DAGM theory. However, it is worth emphasizing that this black hole solution is also a (particular) result of the dimensional regularization suggested in the (controversial) work by Glavan and Lin \cite{PhysRevLett.124.081301}. Henceforth, the black hole metric under consideration can be safely used, since it is considered as a consistent solution \cite{PhysDarkUniverse.31.100748}. Here, we will investigate the test fields, namely, the scalar, electromagnetic, and Dirac fields, which means that we will solve the Klein-Gordon, Maxwell, and Dirac equations in the background under consideration.

In this sense, we can also mention some works on the regularized/consistent 4DEGB gravity, as follows. Motivated by the recent results on searching for the metric solutions of the 4DEGB gravity in the spherically symmetric spacetime, as well as in the Friedmann-Lema\^{i}tre-Robertson-Walker (FLRW) spacetime, it was proved that, in a cylindrically symmetric spacetime, the novel 4DEGB gravity always has the same solutions as the ones in its regularized counterpart. Thus, these solutions satisfy the equations of motion in the novel 4DEGB gravity. In these works, the nontrivial effect of the GB coupling constant on the energy scale of the second order phase transitions (which give rise to cosmic strings) is also shown \cite{EurPhysJC.80.662,EurPhysJC.80.937,EurPhysJC.80.1033,Universe.6.103,PhysRevD.101.104018,EurPhysJPlus.136.436}. In addition, the innermost stable circular orbit and shadow of the 4DEGB black hole were studied by Guo and Li \cite{EurPhysJC.80.588}, where was pointed out that a negative GB coupling constant is allowed to retain a black hole in 4DEGB gravity and gave the allowed range.

In this paper, we calculate the spectrum of quasibound states, and their corresponding radial and angular wave eigenfunctions, for scalar, electromagnetic, and Dirac particles in an asymptotically flat 4-dimensional Einstein--Gauss--Bonnet black hole (4DEGBBH) by using the polynomial condition of the general Heun function. We show that the quasibound states depend on the GB coupling constant, $a$, and that all the test fields constitute stable systems, when $0 < a < 1/2$. In addition, we also compute both radial and angular wave eigenfunctions of the test fields in the 4DEGBBH spacetime.

The paper is organized as follows. In Section \ref{4DEGBBH}, we introduce the metric corresponding to the 4DEGBBH spacetime. In Section \ref{Master_wave}, we solve the master wave equations in the background under consideration. Section \ref{Quasibound_states} is devoted to the quasibound states of the test fields. In Section \ref{Wave_functions}, we provide both radial and angular wave eigenfunctions, by using some properties of the general Heun functions. Finally, in Section \ref{Conclusions}, we summarize the obtained results. Here we adopt the natural units where $G \equiv c \equiv \hbar \equiv 1$.
%
%
\section{The Consistent 4-Dimensional Einstein--Gauss--Bonnet Black Hole Spacetime}\label{4DEGBBH}
A crucial aspect for our studies of black hole radiation (be it an emission, transmission and/or reflection of quasinormal modes and/or quasibound states) is such that the black hole solution should be a solution of the truly 4DAGM theory \cite{PhysLettB.810.135843}, as well as of the theories with extra scalar degrees of freedom \cite{PhysRevD.102.024025,JCAP.07.013,PhysLettB.808.135657,PhysLettB.809.135717}. Thus, we will use the description given by Churilova (see Ref.~\cite{PhysDarkUniverse.31.100748} and references therein), in which all of the aforementioned approaches are taken into account to construct the novel consistent 4DEGB theory, and its black hole solution as well. In what follows, we briefly review the basic ideas behind this approach.

Let us consider a 4-dimensional spacetime in which is valid the following gauge condition \cite{JCAP.09.014,JCAP.05.E01}
\begin{equation}
^{3}\mathcal{G} = \sqrt{\gamma}D_{k}D^{k}(\pi^{ij}\gamma_{ij}/\sqrt{\gamma}) \approx 0,
\label{eq:gauge_condition}
\end{equation}
where $D_{k}$ is the covariant derivative and $\pi^{ij}$ is the canonical momentum conjugate to $\gamma_{ij}$. Then, it can be shown that the unique gravitational action is given by
\begin{equation}
S_{\rm AGM}=\int dt\ d^{3}x\ N\ \sqrt{\gamma}\ \mathcal{L}_{\rm EGB}^{\rm 4D},
\label{eq:action_AGM}
\end{equation}
where $N$ is a lapse function. The 4DEGB Lagrangian $\mathcal{L}_{\rm EGB}^{\rm 4D}$ is defined as
\begin{equation}
\mathcal{L}_{\rm EGB}^{\rm 4D}=\frac{M_{\rm Pl}^{2}}{2}\biggl\{2R-\mathcal{M}+\frac{\tilde{a}}{2}\biggl[8R^{2}-4R\mathcal{M}-\mathcal{M}^{2}-\frac{8}{3}(8R_{ij}R^{ij}-4R_{ij}\mathcal{M}^{ij}-\mathcal{M}_{ij}\mathcal{M}^{ij})\biggr] \biggr\},
\label{eq:4DEGB_Lagrangian}
\end{equation}
where $M_{\rm Pl}$ is the reduced Planck mass characterizing the gravitational coupling strength, $\tilde{a}$ is the GB coupling constant, $R_{ij}$ is the Ricci tensor and $R$ is the Ricci scalar. The $\mathcal{M}_{ij}$ tensor is given by
\begin{equation}
\mathcal{M}_{ij}=R_{ij}+\mathcal{K}_{k}^{k}\mathcal{K}_{ij}-\mathcal{K}_{ik}\mathcal{K}_{j}^{k},
\label{eq:M_tensor}
\end{equation}
with
\begin{equation}
\mathcal{K}_{ij}=(\dot{\gamma}_{ij}-2D_{(i}N_{j)}-\gamma_{ij}D^{2}\lambda_{\rm GF})/2N,
\label{eq:K_EGD}
\end{equation}
where the dot denotes the derivative with respect to the time $t$, and $\lambda_{\rm GF}$ is a gauge-fixing parameter. It worth noticing that this consistent 4DEGB theory has a spatial diffeomorphism invariance and a time reparametrization symmetry given by $t \rightarrow t=t(t')$. Finally, by doing an appropriate rescaling of the GB coupling constant, an exact solution describing the 4DEGBBH spacetime has the following form
\begin{equation}
ds^{2}=-f(r)\ dt^{2}+\frac{1}{f(r)}\ dr^{2}+r^{2}\ d\theta^{2}+r^{2}\sin^{2}\theta\ d\phi^{2},
\label{eq:4DEGBBH_metric}
\end{equation}
where the metric function, $f(r)$, is given by
\begin{equation}
f_{\pm}(r)=1+\frac{r^{2}}{a}\biggl(1 \pm \sqrt{1+\frac{4aM}{r^{3}}}\biggr).
\label{eq:f(r)_4DEGBBH}
\end{equation}
Note that we have chosen the gauge-fixing parameter as $\lambda_{\rm GF}=0$. The parameter $M$ is the total mass centered at the origin of the system of coordinates.

\begin{figure}[h]
\includegraphics[width=1\columnwidth]{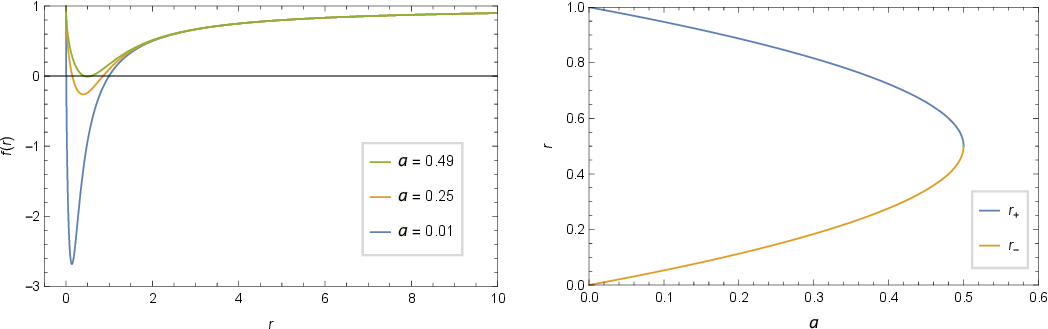}
\caption{Left: the metric function $f(r)$ for $M=1/2$. Right: the event horizons $r_{\pm}$ for $M=1/2$.}
\label{fig:Fig1_4DEGBBH}
\end{figure}

The metric function $f_{+}(r)$ corresponds to an asymptotically de Sitter spacetime, while $f_{-}(r)$ corresponds to an asymptotically flat spacetime. Here, we will focus on the ``minus'' case and hence we will set $f(r) \equiv f_{-}(r)$. In this case, there are two solutions when $a > 0$; otherwise, for $a < 0$ (and $M > 0$) the metric function is not real for sufficiently small values of $r$, that is, for $r^{3} < -4aM$. Here, we will focus on the positive values of the GB coupling constant and hence the surface equation can be parametrized as
\begin{equation}
f(r) = 0 = (r-r_{+})(r-r_{-}).
\label{eq:surface_equation}
\end{equation}
Thus, the two solutions of this surface Equation (\ref{eq:surface_equation}) are the exterior, $r_{+}$, and interior, $r_{-}$, event horizons, and given by
\begin{equation}
r_{\pm}=M \pm \sqrt{M^{2}-\frac{a}{2}}.
\label{eq:event_horizons_4DEGBBH}
\end{equation}
For simplicity, and without loss of generality, from now on we set $M=1/2$, which means that the GB coupling constant is such that  $0 < a < 1/2$; that is the stability region. The behavior of the metric function $f(r)$, as well as the event horizons, is shown in Figure~\ref{fig:Fig1_4DEGBBH}. Thus, the 4DEGBBH metric is well-behaved outside the exterior event horizon, and then we will obtain some results on quasibound states that are valid for these values of $a$.

In what follows, we will solve the equations of motion for the test fields and then discuss the quasibound states, as well as the radial and angular wave eigenfunctions, in the background under consideration.
%
%
\section{Master Wave Equations}\label{Master_wave}
The general covariant equations of motion for the scalar, $\Phi$, electromagnetic, $F_{\mu\nu}$, and Dirac, $\varUpsilon$, massless test fields can be written as
\begin{equation}
\frac{1}{\sqrt{-g}}\partial_{\mu}(g^{\mu\nu}\sqrt{-g}\partial_{\nu}\Phi)=0,
\label{eq:massless_KG}
\end{equation}
\begin{equation}
\frac{1}{\sqrt{-g}}\partial_{\mu}(F^{\mu\nu}\sqrt{-g})=0,
\label{eq:massless_M}
\end{equation}
\begin{equation}
\gamma^{\alpha}\biggl(\frac{\partial}{\partial x^{\alpha}}-\Gamma_{\alpha}\biggl)\varUpsilon=0,
\label{eq:massless_D}
\end{equation}
where $F^{\mu\nu}(=\partial_{\mu}A_{\nu}-\partial_{\nu}A_{\mu})$ is the electromagnetic field tensor, $A_{\mu}$ is the 4-vector electromagnetic potential, $\gamma^{\alpha}$ are the noncommutative gamma matrices, and $\Gamma_{\alpha}$ are the spin connections in the tetrad formalism.

In order to solve these master equations, first we need to separate their angular and radial parts. To do this, we will follow the approaches described by Konoplya et al. \cite{EurPhysJC.80.1049}, Churilova \cite{AnnPhys.427.168425}, and Arag\'{o}n et al. \cite{PhysRevD.103.064006} to deal with the Maxwell, Klein-Gordon, and Dirac equations, respectively. These approaches have the advantage of avoiding square roots of the metric function $f(r)$, so that it will be possible to find analytical solutions for the radial parts. However, these approaches involve too much algebra, so let us summarize it as follows.

For the scalar field, we will use the subscript ``$\mbox{{\tiny B}}$'' (Bosonic, for simplicity but without loss of generality) and set the following ansatz for the wave function:\linebreak $\Phi=u_{\mbox{{\tiny B}}}(r)P^{m}_{\nu}(\cos\theta)\mbox{e}^{im\phi}\mbox{e}^{-i \omega_{\mbox{{\tiny B}}} t}$. Here, $u_{\mbox{{\tiny B}}}(r)=R_{\mbox{{\tiny B}}}(r)/r$ is the scalar radial function, $P^{m}_{\nu}(\cos\theta)$ are the associated Legendre functions, $\omega_{\mbox{{\tiny B}}}$ is the scalar frequency, $m \leq 0$ ($\in \mathbb{Z}$) is the magnetic quantum number, and $\nu$ ($\in \mathbb{C}$) is the degree (or azimuthal quantum number). Thus, the bosonic radial function $R_{\mbox{{\tiny B}}}(r)$ satisfies the following equation,
\begin{equation}
\frac{d^{2} R_{\mbox{{\tiny B}}}(r)}{d r^{2}}+\frac{1}{f(r)}\frac{d f(r)}{d r}\frac{d R_{\mbox{{\tiny B}}}(r)}{d r}+\frac{1}{f^{2}(r)}\biggl\{\omega_{\mbox{{\tiny B}}}^{2}-f(r)\biggl[\frac{\lambda}{r^{2}}+\frac{1}{r}\frac{d f(r)}{d r}\biggr]\biggr\}R_{\mbox{{\tiny B}}}(r)=0,
\label{eq:radial_B_4DEGBBH}
\end{equation}
where $\lambda$ is the separation constant.

For the Maxwell field, we will use the subscript ``$\mbox{{\tiny E}}$'' (Electromagnetic, for simplicity but without loss of generality) and expand the four-vector electromagnetic potential $A_{\mu}$ in terms of the vector associated Legendre functions, as well as assume the time dependence as $\mbox{e}^{-i \omega_{\mbox{{\tiny E}}} t}$, where $\omega_{\mbox{{\tiny E}}}$ is the electromagnetic frequency. Here, $u_{\mbox{{\tiny E}}}(r)=R_{\mbox{{\tiny E}}}(r)/r$ is the electromagnetic radial function. Thus, the electromagnetic radial function $R_{\mbox{{\tiny E}}}(r)$ satisfies the following equation,
\begin{equation}
\frac{d^{2} R_{\mbox{{\tiny E}}}(r)}{d r^{2}}+\frac{1}{f(r)}\frac{d f(r)}{d r}\frac{d R_{\mbox{{\tiny E}}}(r)}{d r}+\frac{1}{f^{2}(r)}\biggl\{\omega_{\mbox{{\tiny E}}}^{2}-f(r)\biggl[\frac{\lambda}{r^{2}}\biggr]\biggr\}R_{\mbox{{\tiny E}}}(r)=0.
\label{eq:radial_E_4DEGBBH}
\end{equation}

For the Dirac field, we will use the subscript ``$\mbox{{\tiny F}}$'' (Fermionic, for simplicity but without loss of generality) and set the following ansatz for the wave function: $\varUpsilon=u_{\pm \mbox{{\tiny F}}}(r)\otimes\varsigma(\theta,\phi)\mbox{e}^{-i \omega_{\mbox{{\tiny F}}} t}$. Here, $u_{\pm \mbox{{\tiny F}}}(r)=R_{\pm \mbox{{\tiny F}}}(r)/rf^{1/4}(r)$ is the fermionic radial function, the signs $\pm$ label the spins, $\omega_{\mbox{{\tiny F}}}$ is the fermionic frequency, and $\varsigma(\theta,\phi)$ is a two-components fermion. Thus, the fermionic radial function $R_{\pm \mbox{{\tiny F}}}(r)$ satisfies the following equation,
\begin{equation}
\frac{d^{2} R_{\pm \mbox{{\tiny F}}}(r)}{d r^{2}}+\biggl[\frac{1}{2f(r)}\frac{d f(r)}{d r}+\frac{1}{r}\biggr]\frac{d R_{\pm \mbox{{\tiny F}}}(r)}{d r}+\frac{1}{f^{2}(r)}\biggl\{\omega_{\mbox{{\tiny F}}}^{2}-f(r)\biggl[\frac{\lambda}{r^{2}}\mp\frac{i\omega_{\mbox{{\tiny F}}}}{r}\pm\frac{i\omega_{\mbox{{\tiny F}}}}{2f(r)}\frac{d f(r)}{d r}\biggr]\biggr\}R_{\pm \mbox{{\tiny F}}}(r)=0.
\label{eq:radial_F_4DEGBBH}
\end{equation}

Indeed, these three radial equations can be condensed in a single one, which is labeled according to the spin, $s$, of the test fields, such that $s=(0,1,1/2)=(\mbox{{\tiny B}},\mbox{{\tiny E}},\mbox{{\tiny F}})$, where we are choosing the ``plus'' spin. Thus, Equations~(\ref{eq:radial_B_4DEGBBH})--(\ref{eq:radial_F_4DEGBBH}) can be rewritten as
\begin{equation}
\frac{d^{2} R_{s}(r)}{d r^{2}}+\biggl[\biggl(\frac{1}{2}\biggr)^{4s(1-s)}\frac{1}{f(r)}\frac{d f(r)}{d r}+\frac{4s(1-s)}{r}\biggr]\frac{d R_{s}(r)}{d r}+\frac{1}{f^{2}(r)}[\omega_{s}^{2}-V_{s}(r,\omega_{s})]R_{s}(r)=0,
\label{eq:radial_4DEGBBH}
\end{equation}
where the effective potential, $V_{s}(r,\omega_{s})$, is given by
\begin{equation}
V_{s}(r,\omega_{s})=f(r)\biggl\{\frac{\lambda}{r^{2}}-\frac{4is(1-s)\omega_{s}}{r}+\biggl[\frac{2(s-1)(s-1/2)}{r}+\frac{4is(1-s)\omega_{s}}{2f(r)}\biggr]\frac{d f(r)}{d r}\biggr\}.
\label{eq:EP_4DEGBBH}
\end{equation}

In fact, the effective potentials $V_{+\frac{1}{2}}$ and $V_{-\frac{1}{2}}$ can be transformed one into another by using the Darboux transformation, which means that both potentials will give the same spectrum of resonant frequencies.

Now, by following the Vieira-Bezerra-Kokkotas (VBK) approach \cite{AnnPhys.373.28,PhysRevD.104.024035}, we can provide an analytical solution for the radial Equation (\ref{eq:radial_4DEGBBH}) in terms of the general Heun functions~\cite{Ronveaux:1995}. It is given by
\begin{equation}
R_{s}^{j}(x) = x^{\frac{1}{2}[\gamma-2^{4s(s-1)}]}\ (x-1)^{\frac{1}{2}[\delta-2^{4s(s-1)}]}\ (x-b)^{\frac{1}{2}[\epsilon-4s(1-s)]}\ [C_{1}^{j}\ y_{1}^{j}(x) + C_{2}^{j}\ y_{2}^{j}(x)],
\label{eq:analytic_solution_radial_4DEGBBH}
\end{equation}
where $C_{1}^{j}$ and $C_{2}^{j}$ are constants to be determined, and $j=\{0,1,b,\infty\}$ labels the solution at each singular point. The new radial coordinate, $x$, and the singularity parameter, $b$, are defined as
\begin{eqnarray}
x & = & \frac{r-r_{-}}{r_{+}-r_{-}},\label{eq:x_4DEGBBH}\\
b & = & -\frac{r_{-}}{r_{+}-r_{-}}.\label{eq:b_4DEGBBH}
\end{eqnarray}
Thus, the pair of linearly independent solutions at $x=0$ ($r=r_{-}$) is given by
\begin{eqnarray}
y_{1}^{0} & = & \mbox{HeunG}(b,q;\alpha,\beta,\gamma,\delta;x),\label{eq:y10}\\
y_{2}^{0} & = & x^{1-\gamma}\mbox{HeunG}(b,(b\delta+\epsilon)(1-\gamma)+q;\alpha+1-\gamma,\beta+1-\gamma,2-\gamma,\delta;x).\label{eq:y20}
\end{eqnarray}
Similarly, the pair of linearly independent solutions corresponding to the exponents $0$ and $1-\delta$ at $x=1$ ($r=r_{+}$) is given by
\begin{eqnarray}
y_{1}^{1} & = & \mbox{HeunG}(1-b,\alpha\beta-q;\alpha,\beta,\delta,\gamma;1-x),\label{eq:y11}\\
y_{2}^{1} & = & (1-x)^{1-\delta}\mbox{HeunG}(1-b,((1-b)\gamma+\epsilon)(1-\delta)+\alpha\beta-q;\alpha+1-\delta,\beta+1-\delta,2-\delta,\gamma;1-x).\label{eq:y21}
\end{eqnarray}
The pair of linearly independent solutions corresponding to the exponents $0$ and $1-\epsilon$ at $x=b$ ($|b|<1$) is given by
\begin{eqnarray}
y_{1}^{b} & = & \mbox{HeunG}\biggl(\frac{b}{b-1},\frac{\alpha\beta b-q}{b-1};\alpha,\beta,\epsilon,\delta;\frac{b-x}{b-1}\biggl),\label{eq:y1x1}\\
y_{2}^{b} & = & \biggl(\frac{b-x}{b-1}\biggl)^{1-\epsilon}\mbox{HeunG}\biggl(\frac{b}{b-1},\frac{(b(\delta+\gamma)-\gamma)(1-\epsilon)}{b-1}+\frac{\alpha\beta b-q}{b-1};\alpha+1-\epsilon,\beta+1-\epsilon,2-\epsilon,\delta;\frac{b-x}{b-1}\biggl).\label{eq:y2x1}
\end{eqnarray}
Finally, the pair of linearly independent solutions corresponding to the exponents $\alpha$ and $\beta$ at $x=\infty$ is given by
\begin{eqnarray}
y_{1}^{\infty} & = & x^{-\alpha}\mbox{HeunG}\biggl(\frac{1}{b},\alpha(\beta-\epsilon)+\frac{\alpha}{b}(\beta-\delta)-\frac{q}{b};\alpha,\alpha-\gamma+1,\alpha-\beta+1,\delta;\frac{1}{x}\biggl),\label{eq:y1i}\\
y_{2}^{\infty} & = & x^{-\beta}\mbox{HeunG}\biggl(\frac{1}{b},\beta(\alpha-\epsilon)+\frac{\beta}{b}(\alpha-\delta)-\frac{q}{b};\beta,\beta-\gamma+1,\beta-\alpha+1,\delta;\frac{1}{x}\biggl).\label{eq:y2i}
\end{eqnarray}
Here, $\mbox{HeunG}(b,q;\alpha,\beta,\gamma,\delta;x)$ denotes a general Heun function, which is analytic in the disk $|x| < 1$, and has a Maclaurin expansion given by
\begin{equation}
\mbox{HeunG}(b,q;\alpha,\beta,\gamma,\delta;x)=\sum_{n=0}^{\infty}c_{n}x^{n},
\label{eq:Maclaurin_HeunG}
\end{equation}
with
\begin{eqnarray}
-qc_{0}+b \gamma c_{1}               & = & 0 \quad (c_{0}=1),\\
P_{n}c_{n-1}-(Q_{n}+q)c_{n}+X_{n}c_{n+1} & = & 0 \quad (n \geq 1),
\label{eq:recursion_HeunG}
\end{eqnarray}
and
\begin{eqnarray}
P_{n} & = & (n-1+\alpha)(n-1+\beta),\nonumber\\
Q_{n} & = & n[(n-1+\gamma)(1+b)+b\delta+\epsilon],\\\nonumber
X_{n} & = & (n+1)(n+\gamma)b.
\label{eq:P_Q_X_recursion_HeunG}
\end{eqnarray}

In these solutions, the parameters $\alpha$, $\beta$, $\gamma$, $\delta$, $\epsilon$, and $q$ are given according to each value of the spin $s$, as follows. For scalar fields, we have
\begin{eqnarray}
\alpha_{\mbox{{\tiny B}}}	& = & \frac{5}{2}-\frac{1}{2}\sqrt{1+\frac{4\lambda}{r_{+}r_{-}}}-\frac{2i\omega_{\mbox{{\tiny B}}}}{r_{+}-r_{-}},\label{eq:alphaB_4DEGBBH}\\
\beta_{\mbox{{\tiny B}}}		& = & -\frac{1}{2}-\frac{1}{2}\sqrt{1+\frac{4\lambda}{r_{+}r_{-}}}-\frac{2i\omega_{\mbox{{\tiny B}}}}{r_{+}-r_{-}},\label{eq:betaB_4DEGBBH}\\
\gamma_{\mbox{{\tiny B}}}	& = & 1-\frac{2i\omega_{\mbox{{\tiny B}}}}{r_{+}-r_{-}},\label{eq:gammaB_4DEGBBH}\\
\delta_{\mbox{{\tiny B}}}	& = & 1-\frac{2i\omega_{\mbox{{\tiny B}}}}{r_{+}-r_{-}},\label{eq:deltaB_4DEGBBH}\\
\epsilon_{\mbox{{\tiny B}}}	& = & 1-\sqrt{1+\frac{4\lambda}{r_{+}r_{-}}},\label{eq:epsilonB_4DEGBBH}\\
q_{\mbox{{\tiny B}}}			& = & \frac{1}{2 r_{-} (r_{+}-r_{-})^4}\biggl\{2 i \omega_{\mbox{{\tiny B}}}[\sqrt{r_{+}^5 r_{-} (4 \lambda +r_{+} r_{-})}+3 \sqrt{r_{+} r_{-}^5 (4 \lambda +r_{+} r_{-})}]\nonumber\\
&& +r_{-} [\sqrt{r_{+}^5 r_{-} (4 \lambda +r_{+} r_{-})}+3 \sqrt{r_{+} r_{-}^5 (4 \lambda +r_{+} r_{-})}]\nonumber\\
&& -r_{+} \biggl\{\sqrt{r_{+}^5 r_{-} (4 \lambda +r_{+} r_{-})}+3 \sqrt{r_{+} r_{-}^5 (4 \lambda +r_{+} r_{-})}-r_{-}^4 \biggl(\sqrt{1+\frac{4 \lambda }{r_{+} r_{-}}}+4\biggr)\nonumber\\
&& +r_{-}^2 [3 \sqrt{r_{+} r_{-} (4 \lambda +r_{+} r_{-})}-8 \omega_{\mbox{{\tiny B}}} ^2-6 \lambda]+6 i r_{-} \omega_{\mbox{{\tiny B}}}  \sqrt{r_{+} r_{-} (4 \lambda +r_{+} r_{-})}+14 i r_{-}^3 \omega_{\mbox{{\tiny B}}} \biggr\}\nonumber\\
&& -r_{+}^4 r_{-}+2 r_{+}^3 (\lambda +2 r_{-}^2-i r_{-} \omega_{\mbox{{\tiny B}}})+r_{+}^2 r_{-} [3 \sqrt{r_{+} r_{-} (4 \lambda +r_{+} r_{-})}-6 r_{-}^2+10 i r_{-} \omega_{\mbox{{\tiny B}}}-6 \lambda]\nonumber\\
&& -r_{-}^5 \biggl(\sqrt{1+\frac{4 \lambda }{r_{+} r_{-}}}+1\biggr)-2 i r_{-}^4 \omega_{\mbox{{\tiny B}}}  \biggl(\sqrt{1+\frac{4 \lambda }{r_{+} r_{-}}}-3\biggr)-2 r_{-}^3 (\lambda +4 \omega_{\mbox{{\tiny B}}} ^2)\biggr\}.\label{eq:qB_4DEGBBH}
\end{eqnarray}
For electromagnetic fields, we have
\begin{eqnarray}
\alpha_{\mbox{{\tiny E}}}	& = & \frac{3}{2}-\frac{1}{2}\sqrt{1+\frac{4\lambda}{r_{+}r_{-}}}-\frac{2i\omega_{\mbox{{\tiny E}}}}{r_{+}-r_{-}},\label{eq:alphaE_4DEGBBH}\\
\beta_{\mbox{{\tiny E}}}		& = & \frac{1}{2}-\frac{1}{2}\sqrt{1+\frac{4\lambda}{r_{+}r_{-}}}-\frac{2i\omega_{\mbox{{\tiny E}}}}{r_{+}-r_{-}},\label{eq:betaE_4DEGBBH}\\
\gamma_{\mbox{{\tiny E}}}	& = & 1-\frac{2i\omega_{\mbox{{\tiny E}}}}{r_{+}-r_{-}},\label{eq:gammaE_4DEGBBH}\\
\delta_{\mbox{{\tiny E}}}	& = & 1-\frac{2i\omega_{\mbox{{\tiny E}}}}{r_{+}-r_{-}},\label{eq:deltaE_4DEGBBH}\\
\epsilon_{\mbox{{\tiny E}}}	& = & 1-\sqrt{1+\frac{4\lambda}{r_{+}r_{-}}},\label{eq:epsilonE_4DEGBBH}\\
q_{\mbox{{\tiny E}}}			& = & \frac{1}{2 r_{-} (r_{+}-r_{-})^4}\biggl\{2 i \omega_{\mbox{{\tiny E}}}[\sqrt{r_{+}^5 r_{-} (4 \lambda +r_{+} r_{-})}+3 \sqrt{r_{+} r_{-}^5 (4 \lambda +r_{+} r_{-})}]\nonumber\\
&& +r_{-} [\sqrt{r_{+}^5 r_{-} (4 \lambda +r_{+} r_{-})}+3 \sqrt{r_{+} r_{-}^5 (4 \lambda +r_{+} r_{-})}]\nonumber\\
&& -r_{+} \biggl\{\sqrt{r_{+}^5 r_{-} (4 \lambda +r_{+} r_{-})}+3 \sqrt{r_{+} r_{-}^5 (4 \lambda +r_{+} r_{-})}+r_{-}^4 \biggl(4-\sqrt{1+\frac{4 \lambda }{r_{+} r_{-}}}\biggr)\nonumber\\
&& +r_{-}^2 [3 \sqrt{r_{+} r_{-} (4 \lambda +r_{+} r_{-})}-8 \omega_{\mbox{{\tiny E}}} ^2-6 \lambda]+6 i r_{-} \omega_{\mbox{{\tiny E}}}  \sqrt{r_{+} r_{-} (4 \lambda +r_{+} r_{-})}+14 i r_{-}^3 \omega_{\mbox{{\tiny E}}} \biggr\}\nonumber\\
&& +r_{+}^4 r_{-}+2 r_{+}^3 (\lambda -2 r_{-}^2-i r_{-} \omega_{\mbox{{\tiny E}}})+r_{+}^2 r_{-} [3 \sqrt{r_{+} r_{-} (4 \lambda +r_{+} r_{-})}+6 r_{-}^2+10 i r_{-} \omega_{\mbox{{\tiny E}}}-6 \lambda]\nonumber\\
&& r_{-}^5 \biggl(\sqrt{1-\frac{4 \lambda }{r_{+} r_{-}}}+1\biggr)-2 i r_{-}^4 \omega_{\mbox{{\tiny E}}}  \biggl(\sqrt{1+\frac{4 \lambda }{r_{+} r_{-}}}-3\biggr)-2 r_{-}^3 (\lambda +4 \omega_{\mbox{{\tiny E}}} ^2)\biggr\}.\label{eq:qE_4DEGBBH}
\end{eqnarray}
Finally, for Dirac fields, we have
\begin{eqnarray}
\alpha_{\mbox{{\tiny F}}}	& = & \frac{3}{2}-\sqrt{\frac{\lambda}{r_{+}r_{-}}}-\frac{2i\omega_{\mbox{{\tiny F}}}}{r_{+}-r_{-}},\label{eq:alphaF_4DEGBBH}\\
\beta_{\mbox{{\tiny F}}}		& = & \frac{1}{2}-\sqrt{\frac{\lambda}{r_{+}r_{-}}}-\frac{2i\omega_{\mbox{{\tiny F}}}}{r_{+}-r_{-}},\label{eq:betaF_4DEGBBH}\\
\gamma_{\mbox{{\tiny F}}}	& = & \frac{3}{2}-\frac{2i\omega_{\mbox{{\tiny F}}}}{r_{+}-r_{-}},\label{eq:gammaF_4DEGBBH}\\
\delta_{\mbox{{\tiny F}}}	& = & \frac{1}{2}-\frac{2i\omega_{\mbox{{\tiny F}}}}{r_{+}-r_{-}},\label{eq:deltaF_4DEGBBH}\\
\epsilon_{\mbox{{\tiny F}}}	& = & 1-2\sqrt{\frac{\lambda}{r_{+}r_{-}}},\label{eq:epsilonF_4DEGBBH}\\
q_{\mbox{{\tiny F}}}			& = & \frac{1}{4 r_{-} (r_{+}-r_{-})^4}\biggl\{24 \sqrt{\lambda  r_{+}^5 r_{-}^3}+8 i \omega_{\mbox{{\tiny F}}}  \sqrt{\lambda  r_{+}^5 r_{-}}+2 r_{+}^4 r_{-}-36 \sqrt{\lambda  r_{+}^3 r_{-}^5}-24 i \omega_{\mbox{{\tiny F}}}  \sqrt{\lambda  r_{+}^3 r_{-}^3}-6 \sqrt{\lambda  r_{+}^7 r_{-}}\nonumber\\
&& +r_{+}^3(4 \lambda -9 r_{-}^2-8 i r_{-} \omega_{\mbox{{\tiny F}}})+r_{+}^2 r_{-} \left(15 r_{-}^2+32 i r_{-} \omega_{\mbox{{\tiny F}}}-12 \lambda \right)-6 \sqrt{\frac{\lambda  r_{-}^9}{r_{+}}}-8 i \omega_{\mbox{{\tiny F}}}  \sqrt{\frac{\lambda  r_{-}^7}{r_{+}}}+24 \sqrt{\lambda  r_{+} r_{-}^7}\nonumber\\
&& +24 i \omega_{\mbox{{\tiny F}}}  \sqrt{\lambda  r_{+} r_{-}^5}+r_{+} r_{-}^2 (12 \lambda -11 r_{-}^2-40 i r_{-} \omega_{\mbox{{\tiny F}}} +16 \omega_{\mbox{{\tiny F}}} ^2)+3 r_{-}^5+16 i r_{-}^4 \omega_{\mbox{{\tiny F}}} -4 r_{-}^3 (\lambda +4 \omega_{\mbox{{\tiny F}}} ^2)\biggr\}.\label{eq:qF_4DEGBBH}
\end{eqnarray}

Next, we will use these analytical solutions of the radial equations in the 4DEGBBH spacetime, and the properties of the general Heun functions, to compute the spectrum of resonant frequencies related to the quasibound states.
%
%
\section{Quasibound States}\label{Quasibound_states}
The quasibound states, also known as quasistationary levels or resonance spectra, are a kind of wave phenomena occurring near the black hole exterior event horizon, that is, they are localized in the black hole potential well. The definition of quasibound states means that there exist a flux of particles crossing into the black hole, so that the spectrum of quasibound states is constituted by complex frequencies, which can be expressed as $\omega=\omega_{R}+i\omega_{I}$, where $\omega_{R}$ and $\omega_{I}$ are the real and imaginary parts, respectively. By using these results, it is possible, in principle, to get some information about the physics of black holes as well as to validate some alternative/modified theories of gravity.

The real part of the resonant frequency is the oscillation frequency, while the imaginary part determines the stability of the system. Thus, the wave solution is said to be stable when the imaginary part of the resonant frequency is negative ($\omega_{I} < 0$), which means a decay rate with the time. Otherwise, the wave solution is unstable when the imaginary part of the resonant frequency is positive ($\omega_{I} > 0$), which means a growth rate with the time.

In order to compute the spectrum of quasibound states, we need to impose two boundary conditions, which are related to the asymptotic behavior of the radial solution. First, it should describe an ingoing wave at the exterior event horizon. Second, it should tend to zero far from the black hole at asymptotic infinity, that is, the probability to find such a particle in the spatial infinity must be zero.

To derive the characteristic resonance equation, many authors have been using a method which consists of solving the radial equation in two different asymptotic regions and then matching these two radial solutions in their common overlap region. However, in this work, we will use the VBK approach (for details, see Refs. \cite{PhysRevD.104.024035,AnnPhys.373.28}) to derive the characteristic resonance equation and then find the spectrum of resonant frequencies related to quasibound states. This approach consists in (i) obtaining the ingoing wave solution at the exterior event horizon, (ii) imposing the polynomial condition of the general Heun functions, and then (iii) computing the resonant frequencies.

Thus, in the limit when $r \rightarrow r_{+}$, which implies that $x \rightarrow 1$, the radial solution given by Equation~(\ref{eq:analytic_solution_radial_4DEGBBH}) behaves as
\begin{equation}
\lim_{r \rightarrow r_{+}} R_{s}^{1}(r) \sim C_{1}^{1}\ (r-r_{+})^{\frac{1}{2}[\delta-2^{4s(s-1)}]} + C_{2}^{1}\ (r-r_{+})^{-\frac{1}{2}[\delta-2^{4s(s-1)}]},
\label{eq:asymptotic_4DEGBBH}
\end{equation}
where $C_{1}^{1}$ and $C_{2}^{1}$ include all remaining constants. On the other hand, from Equations~(\ref{eq:deltaB_4DEGBBH}), (\ref{eq:deltaE_4DEGBBH}) and (\ref{eq:deltaF_4DEGBBH}), we get
\begin{equation}
\frac{1}{2}[\delta-2^{4s(s-1)}]=-\frac{i\omega_{s}}{2\kappa_{+}},
\label{eq:D0_QBSs_full_4DEGBBH}
\end{equation}
where $\kappa_{+}$ is the gravitational acceleration on the exterior event horizon and given by
\begin{equation}
\kappa_{+} \equiv \frac{1}{2} \left.\frac{d f(r)}{dr}\right|_{r=r_{+}} = \frac{r_{+}-r_{-}}{2}.
\label{eq:grav_acc_4DEGBBH}
\end{equation}
Then, we can rewrite Equation~(\ref{eq:asymptotic_4DEGBBH}) as
\begin{equation}
\lim_{r \rightarrow r_{+}} R_{s}^{1}(r) \sim C_{1}^{1}\ R_{\rm in}^{1}(r) + C_{2}^{1}\ R_{\rm out}^{1}(r),
\label{eq:full_wave_4DEGBBH}
\end{equation}
where the ingoing, $R_{\rm in}^{1}(r)$, and outgoing, $R_{\rm out}^{1}(r)$, wave solutions are given by
\begin{eqnarray}
R_{\rm in}^{1}(r>r_{+})		& = & (r-r_{+})^{-\frac{i\omega_{s}}{2\kappa_{+}}},\label{eq:radial_in_4DEGBBH}\\
R_{\rm out}^{1}(r>r_{+})	& = & (r-r_{+})^{\frac{i\omega_{s}}{2\kappa_{+}}}.\label{eq:radial_out_4DEGBBH}
\end{eqnarray}
Therefore, in order to fully satisfy the first boundary condition, we must impose that $C_{2}^{1}=0$ in Equation~(\ref{eq:full_wave_4DEGBBH}), as well as in Equation~(\ref{eq:analytic_solution_radial_4DEGBBH}). Finally, we obtain the radial wave solution describing ingoing test particles at the 4DEGBBH exterior event horizon, which is given by
\begin{equation}
\lim_{r \rightarrow r_{+}} R_{s}^{1}(r) \sim C_{1}^{1}\ R_{\rm in}^{1}(r>r_{+}).
\label{eq:full_wave_2_4DEGBBH}
\end{equation}

On the other hand, in the limit when $r \rightarrow \infty$, which implies that $x \rightarrow \infty$, the radial solution given by Equation~(\ref{eq:analytic_solution_radial_4DEGBBH}) behaves as
\begin{equation}
\lim_{r \rightarrow \infty} R_{s}^{\infty}(r) \sim C_{1}^{\infty}\ r^{\sigma_{s}},
\label{eq:infty_solution_radial_2_4DEGBBH}
\end{equation}
with $\sigma_{s}=D_{s}-\alpha_{s}$, where the coefficients $D_{s}$ are given by
\begin{eqnarray}
D_{\mbox{{\tiny B}}}	& = & -\frac{1}{2}-\frac{1}{2}\sqrt{1+\frac{4\lambda}{r_{+}r_{-}}}-\frac{2i\omega_{\mbox{{\tiny B}}}}{r_{+}-r_{-}},\label{eq:DB_4DEGBBH}\\
D_{\mbox{{\tiny E}}}	& = & -\frac{1}{2}-\frac{1}{2}\sqrt{1+\frac{4\lambda}{r_{+}r_{-}}}-\frac{2i\omega_{\mbox{{\tiny E}}}}{r_{+}-r_{-}},\label{eq:DE_4DEGBBH}\\
D_{\mbox{{\tiny F}}}	& = & -1+\frac{1}{r_{+}-r_{-}}\biggl(\sqrt{\frac{r_{-}\lambda}{r_{+}}}-\sqrt{\frac{r_{+}\lambda}{r_{-}}}-2i\omega_{\mbox{{\tiny F}}}\biggr).\label{eq:DF_4DEGBBH}
\end{eqnarray}
Thus, the sign of the real part of $\sigma_{s}$ determines the asymptotic behavior of the radial solution far from the black hole at asymptotic infinity. Therefore, the radial solution tends to zero if $\mbox{Re}[\sigma_{s}] < 0$ and then it will fully satisfy the second boundary condition, which describes the quasibound states. Otherwise, if $\mbox{Re}[\sigma_{s}] > 0$, the radial solution diverges. The final asymptotic behavior of the radial solution will be determined when we know the values of the frequencies $\omega_{s}$.

Now, to find the characteristic resonance equation, we match these two asymptotic radial solutions in their common overlap region by imposing the polynomial conditions of the general Heun functions, which are given by
\begin{eqnarray}
\alpha		& = & -n,\label{eq:alpha-condition}\\
c_{n+1}(q)	& = & 0,\label{eq:q-condition}
\end{eqnarray}
where $n=0,1,2,\ldots$ is the overtone number, which can be, without loss of generality, called the principal quantum number. From the first polynomial condition, given by Equation~(\ref{eq:alpha-condition}), we can find the frequency eigenvalues. On the other hand, from the second polynomial condition, given by Equation~(\ref{eq:q-condition}), we can determine the values of the separation constant $\lambda$ for each value (mode) $n$, and then write the radial and angular wave eigenfunctions. Thus, by imposing Equation~(\ref{eq:alpha-condition}) on each parameter $\alpha_{s}$, we obtain the exact analytical spectrum of quasibound states for massless test fields in the 4DEGBBH spacetime. They are given by
\begin{eqnarray}
\omega_{\mbox{{\tiny B}}n}	& = & -\frac{1}{4}i\sqrt{1-2a}\biggl(5+2n-\sqrt{\frac{a+8\lambda}{a}}\biggr),\label{eq:omegaB_4DEGBBH}\\
\omega_{\mbox{{\tiny E}}n}	& = & -\frac{1}{4}i\sqrt{1-2a}\biggl(3+2n-\sqrt{\frac{a+8\lambda}{a}}\biggr),\label{eq:omegaE_4DEGBBH}\\
\omega_{\mbox{{\tiny F}}n}	& = & -\frac{1}{4}i\sqrt{1-2a}\biggl(3+2n-\sqrt{\frac{8\lambda}{a}}\biggr).\label{eq:omegaF_4DEGBBH}
\end{eqnarray}
On the other hand, by imposing Equation~(\ref{eq:q-condition}) on each parameter $q_{s}$, we determine the exact analytical value of the separation constant $\lambda_{sn}$ (for details, see Refs. \cite{EurPhysJC.81.1143,PhysRevD.105.045015,EurPhysJC.82.669}). However, the final expressions are quite long, and for this reason no insight is gained by writing them out. Thus, instead of doing this, we will present some of their features in what follows.

\begin{figure}[h]
\includegraphics[scale=1]{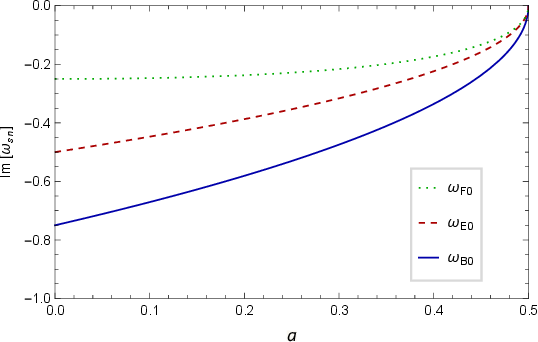}
\caption{The quasibound states $\omega_{sn}$ in the 4DEGBBH spacetime. The plot shows the decay rate $\mbox{Im}[\omega_{sn}]$ for the fundamental mode $n=0$, as a function of the GB coupling constant $a$.}
\label{fig:Fig2_4DEGBBH}
\end{figure}

In Table \ref{tab:I_4DEGBBH}, we present some values of the quasibound state $\omega_{sn}$, as well as the corresponding separation constants $\lambda_{sn}$ and coefficients $\sigma_{sn}=D_{sn}+n$, as functions of the GB coupling constant $a$. We also show the behavior of the quasibound states $\omega_{sn}$ in Figure~\ref{fig:Fig2_4DEGBBH}, as functions of the GB coupling constant $a$. From these results, we can conclude that all solution for the massless resonant frequencies $\omega_{sn}$ are physically admissible in the fundamental mode, which represents the spectrum of quasibound states in the 4DEGBBH spacetime. In this scenario, the radial solution given by Equation~(\ref{eq:analytic_solution_radial_4DEGBBH}) tends to zero far from the 4DEGBBHs at asymptotic infinity, since $\mbox{Re}[\sigma_{sn}] < 0$. In addition, we can see that the motion of all the test fields are overdamped (with purely imaginary frequencies). Furthermore, all the test fields may describe stable systems, since there is no change in the sign of the imaginary part of their massless resonant frequencies in the stability region, that is, when $0 < a < 1/2$. We can also see that the imaginary part of the massless resonant frequencies $\omega_{sn}$ (for all the test fields in these modes) decreases (in modulus) as the GB coupling constant $a$ approaches to 1/2.

It is worth noticing that we cannot determine the quasibound states $\omega_{sn}$, nor the separation constants $\lambda_{sn}$ and the coefficients $\sigma_{sn}$, for the particular case when $a=0$, which corresponds to the General Relativity (GR) limit. This is due to the fact that when we set $a=0$, the singularity parameter $b$ goes to 0 and hence the radial solution must be recalculated in this scenario, since it will occur a confluent process involving the already existing singularity at the point $x=0$ with the ones in the point $x=b \rightarrow 0$, which will lead to a new radial solution given in terms of the confluent Heun functions, with totally new parameters. Henceforth, for the case of scalar fields, we can compare this result with the one obtained by Muniz et al. \cite{JCAP.01.006}, where the spectrum of quasibound states for the Schwarzschild black hole ($a=0$, standard GR case) is such that $\omega_{n}=-i(n+1)/2$, for $M=1/2$; we see that there exist a peculiar difference between the found solutions.

\begin{figure}[h]
\includegraphics[scale=1]{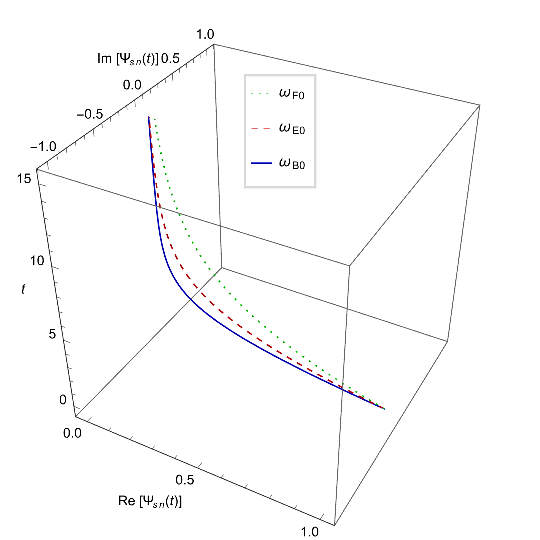}
\caption{The time dependence of the test fields in the 4DEGBBH spacetime for $a=0.15$. We focus on the fundamental mode $n=0$.}
\label{fig:Fig3_4DEGBBH}
\end{figure}

Finally, we can also discuss the time dependence of the wave eigenfunctions, $\Psi_{sn}(t)=\mbox{e}^{-i\omega_{sn}t}$, which is shown in Figure~\ref{fig:Fig3_4DEGBBH}. From this plot, we can realize the ``final flight'' of the test particles when crossing into the 4DEGBBH exterior event horizon.

\begin{table}[h]
\caption{The quasibound states $\omega_{sn}$, the separation constants $\lambda_{sn}$, and the real part of coefficients $\sigma_{sn;\eta}$. We focus on the fundamental mode $n=0$.}
\label{tab:I_4DEGBBH}


\begin{tabular}{cccccccccc}
\toprule\noalign{\smallskip}
\boldmath{$a$}    & \boldmath{\textbf{$\omega_{\mbox{{\tiny B}}0}$}} & \boldmath{\textbf{$\lambda_{\mbox{{\tiny B}}0;+}$}} & \boldmath{\textbf{$\mbox{Re}[\sigma_{\mbox{{\tiny B}}0}]$}} & \boldmath{\textbf{$\omega_{\mbox{{\tiny E}}0}$}} & \boldmath{\textbf{$\lambda_{\mbox{{\tiny E}}0;+}$}} & \boldmath{\textbf{$\mbox{Re}[\sigma_{\mbox{{\tiny E}}0}]$}} & \boldmath{\textbf{$\omega_{\mbox{{\tiny F}}0}$}} & \boldmath{\textbf{$\lambda_{\mbox{{\tiny F}}0;+}$}} & \boldmath{\textbf{$\mbox{Re}[\sigma_{\mbox{{\tiny F}}0}]$}} \\
\noalign{\smallskip}\hline
$0.05$ & $-0.711512i$ & $0.018750$ & $-3$ & $-0.474342i$ & $0$ & $-2$ & $-0.249342i$ & $0.023734$ & $-2.500000$ \\
$0.10$ & $-0.670820i$ & $0.037500$ & $-3$ & $-0.447214i$ & $0$ & $-2$ & $-0.247214i$ & $0.044861$ & $-2.500000$ \\
$0.15$ & $-0.627495i$ & $0.056250$ & $-3$ & $-0.418330i$ & $0$ & $-2$ & $-0.243330i$ & $0.063250$ & $-2.500000$ \\
$0.20$ & $-0.580948i$ & $0.075000$ & $-3$ & $-0.387298i$ & $0$ & $-2$ & $-0.237298i$ & $0.078730$ & $-2.500000$ \\
$0.25$ & $-0.530330i$ & $0.093750$ & $-3$ & $-0.353553i$ & $0$ & $-2$ & $-0.228553i$ & $0.091069$ & $-2.500000$ \\
$0.30$ & $-0.474342i$ & $0.112500$ & $-3$ & $-0.316228i$ & $0$ & $-2$ & $-0.216228i$ & $0.099934$ & $-2.500000$ \\
$0.35$ & $-0.410792i$ & $0.131250$ & $-3$ & $-0.273861i$ & $0$ & $-2$ & $-0.198861i$ & $0.104801$ & $-2.500000$ \\
$0.40$ & $-0.335410i$ & $0.150000$ & $-3$ & $-0.223607i$ & $0$ & $-2$ & $-0.173607i$ & $0.104721$ & $-2.500000$ \\
$0.45$ & $-0.237171i$ & $0.168750$ & $-3$ & $-0.158114i$ & $0$ & $-2$ & $-0.133114i$ & $0.097451$ & $-2.500000$ \\
\noalign{\smallskip}\hline
\end{tabular}
\end{table}

%
%
\subsection*{Transition Frequency}\label{Transition_frequency}
In addition, we can calculate the transition frequency, $\Delta \omega$, between two highly damped ($n \rightarrow \infty$) neighboring states \cite{PhysRevD.94.084040}. It is given by
\begin{equation}
\Delta \omega \approx \mbox{Im}[\omega_{s(n-1)}]-\mbox{Im}[\omega_{sn}] = \kappa_{+} = \frac{1}{2}\sqrt{1-2a}.
\label{eq:transition_4DEGBBH}
\end{equation}
On the other hand, the natural adiabatic invariant quantity, $I_{\rm ad}$, for a test field-black hole system with total energy $E$, is given by
\begin{equation}
I_{\rm ad}=\int\frac{dE}{\Delta \omega} = \int\frac{T_{\rm H}dS_{\rm BH}}{\Delta \omega} = \frac{\hbar S_{\rm BH}}{2\pi},
\label{eq:adiabatic_invariant_4DEGBBH}
\end{equation}
where $T_{\rm H}$ is the Hawking temperature and $S_{\rm BH}(=\mathcal{A}_{+}/4\hbar)$ is the Bekenstein-Hawking entropy, with $\mathcal{A}_{+}$ being the surface area of the exterior event horizon. In this limit, the Bohr-Sommerfeld quantization condition applies and hence $I_{\rm ad}$ is a quantized quantity, namely, $I_{\rm ad}=n\hbar$. Thus, from Equation~(\ref{eq:adiabatic_invariant_4DEGBBH}) we get
\begin{equation}
S_{{\rm BH}n}=2 \pi n.
\label{eq:entropy_4DEGBBH}
\end{equation}
Therefore, the area spectrum is given by
\begin{equation}
\mathcal{A}_{+n} = 8 \pi n \hbar,
\label{eq:area_spectrum_4DEGBBH}
\end{equation}
so that its minimum change becomes
\begin{equation}
\Delta \mathcal{A}^{\rm min}_{+} = 8 \pi \hbar.
\label{eq:minimum_4DEGBBH}
\end{equation}
From Equations~(\ref{eq:entropy_4DEGBBH}) and (\ref{eq:area_spectrum_4DEGBBH}), we can conclude that both entropy and area spectra are equally spaced. Furthermore, they do not depend on the black hole parameters. Finally, Equation~(\ref{eq:minimum_4DEGBBH}) shows that the 4DEGBBH exterior event horizon is made by patches with equal area.

Next, we will use the polynomial conditions of the general Heun functions, as well as the frequency eigenvalues, to discuss both radial and angular wave eigenfunctions.
%
%
\section{Wave Eigenfunctions}\label{Wave_functions}
In this section, we continue to use the VBK approach (for details, see Refs. \cite{EurPhysJC.81.1143,PhysRevD.105.045015,EurPhysJC.82.669}) in order to present both radial and angular wave eigenfunctions describing the quasibound states of massless test particles propagating in the 4DEGBBH spacetime.
%
%
\subsection{Radial Eigenfunctions}\label{Radial_eigenfunctions}
The quasibound state radial wave eigenfunctions are given in terms of the general Heun polynomials, which can be denoted as $\mbox{HeunGp}_{n;\eta}(b,q;-n,\beta,\gamma,\delta;x)$, where the parameter $\eta$ is related to the appropriate determination of the accessory parameter $q=q_{n;\eta}$ via the second polynomial condition given by Equation~(\ref{eq:q-condition}), which are the solutions that cut (in a certain order) the power series describing the general Heun functions; it is such that $\eta=0,1,2,\ldots,n$. In our case, the general Heun polynomials depend also on the spin $s$ and hence they should be denoted as $\mbox{HeunGp}_{sn;\eta}(b,q_{sn;\eta};-n,\beta,\gamma,\delta;x)$.

Therefore, the quasibound state radial wave eigenfunctions for massless test particles propagating in the 4DEGBBH spacetime are given by
\vspace{-12pt}
\begin{equation}
R_{sn;\eta}(x) = C_{sn;\eta}\ x^{\frac{1}{2}[\gamma-2^{4s(s-1)}]}\ (x-1)^{\frac{1}{2}[\delta-2^{4s(s-1)}]}\ (x-b)^{\frac{1}{2}[\epsilon-4s(1-s)]}\ \mbox{HeunGp}_{sn;\eta}(b,q_{sn;\eta};-n,\beta,\gamma,\delta;x),
\label{eq:eigenfunctions_4DEGBBH}
\end{equation}
where $C_{sn;\eta}$ is a constant to be determined. Thus, the full quasibound state radial wave eigenfunctions, $u_{sn;\eta}(r)$, are given by
\begin{eqnarray}
u_{\mbox{{\tiny B}}n;\eta}(r) & = & \frac{R_{\mbox{{\tiny B}}n;\eta}(r)}{r},\label{eq:function_uB_4DEGBBH}\\
u_{\mbox{{\tiny E}}n;\eta}(r) & = & \frac{R_{\mbox{{\tiny E}}n;\eta}(r)}{r},\label{eq:function_uE_4DEGBBH}\\
u_{\mbox{{\tiny F}}n;\eta}(r) & = & \frac{R_{\mbox{{\tiny F}}n;\eta}(r)}{rf^{1/4}(r)}.\label{eq:function_uF_4DEGBBH}
\end{eqnarray}

\begin{figure}[h]
\includegraphics[width=1\columnwidth]{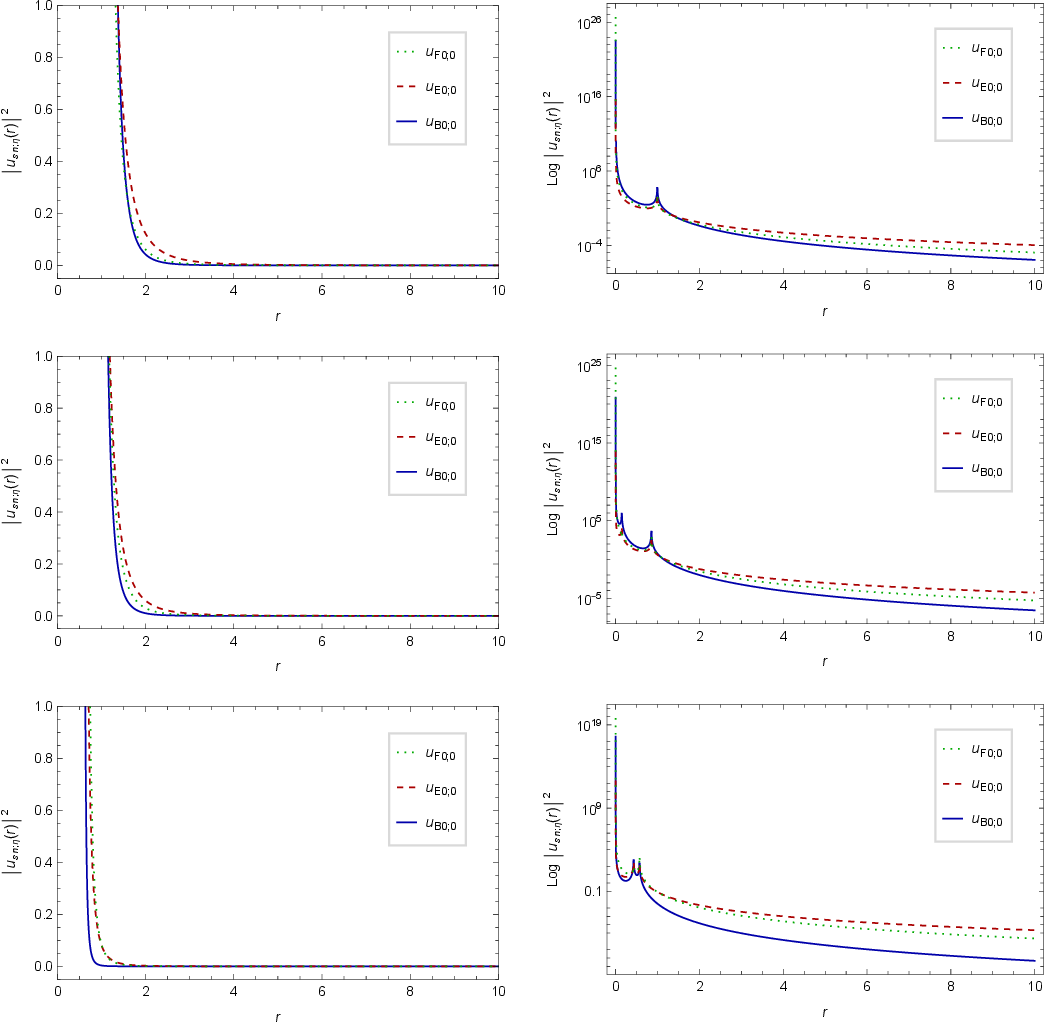}
\caption{\textbf{Top} panel: the squared full quasibound state radial wave eigenfunctions (\textbf{left}) and their log-scale plot (\textbf{right}) for $a=0.01$. \textbf{Middle} panel: the squared full quasibound state radial wave eigenfunctions (\textbf{left}) and their log-scale plot (\textbf{right}) for $a=0.25$. \textbf{Bottom} panel: the squared full quasibound state radial wave eigenfunctions (\textbf{left}) and their log-scale plot (\textbf{right}) for $a=0.49$. We focus on the fundamental mode $n=0$, and the units are in multiples of $C_{sn;\eta}$.}
\label{fig:Fig4_4DEGBBH}
\end{figure}

The squared full quasibound state radial wave eigenfunctions are presented in \mbox{Figure~\ref{fig:Fig4_4DEGBBH}}, as functions of the radial coordinate $r$ for some values of the GB coupling constant $a$. From these results, we can conclude that all the resonant frequencies $\omega_{sn}$, given by \mbox{Equations~(\ref{eq:omegaB_4DEGBBH})--(\ref{eq:omegaF_4DEGBBH})}, describe massless quasibound states in the 4DEGBBH spacetime, since the radial solutions tend to zero at asymptotic infinity and diverge at the exterior event horizon. Note that these eigenfunctions reach a maximum value (at the exterior event horizon $r_{+}=0.994975$ for $a=0.01$, $r_{+}=0.853553$ for $a=0.25$, and $r_{+}=0.570711$ for $a=0.49$) and then cross into the 4DEGBBH, as shown in the log-scale plots.

%
%
\subsection{Angular Eigenfunctions}\label{Angular_eigenfunctions}
It is known that the angular solution must be regular at its two boundaries, namely, when $\theta=0$ and $\theta=\pi$. This requirements single out a discrete set of angular eigenvalues $\lambda$, which couples the angular and radial equations. In our case, we can obtain the values for the angular eigenvalues $\lambda_{sn;\pm}$ by substituting the first polynomial condition, given by Equation~(\ref{eq:alpha-condition}), into the second polynomial condition, given by Equation~(\ref{eq:q-condition}). The $\pm$signs denote the positive and negative solutions; Here, we choose the positive values, in order to express all the angular solutions in terms of the associated Legendre functions.

\begin{figure}[h]
\includegraphics[scale=0.6]{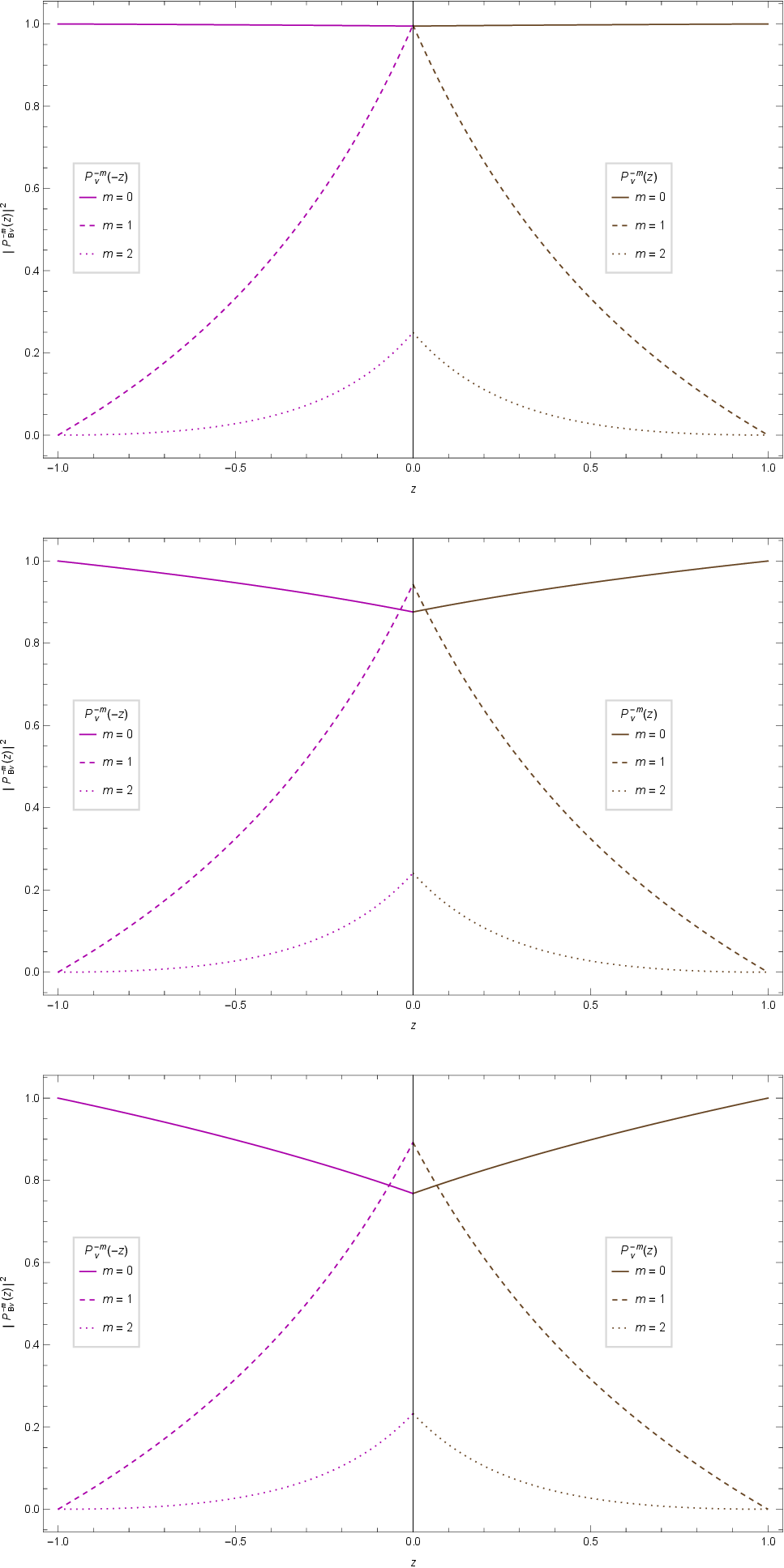}
\caption{\textbf{Top} panel: the first three squared scalar quasibound state angular wave eigenfunctions for $a=0.01$. \textbf{Middle} panel: the first three squared scalar quasibound state angular wave eigenfunctions for $a=0.25$. \textbf{Bottom} panel: the first three scalar squared quasibound state angular wave eigenfunctions for $a=0.49$. We focus on the fundamental mode $n=0$.}
\label{fig:Fig5_4DEGBBH}
\end{figure}

\begin{figure}[h]
\includegraphics[scale=0.65]{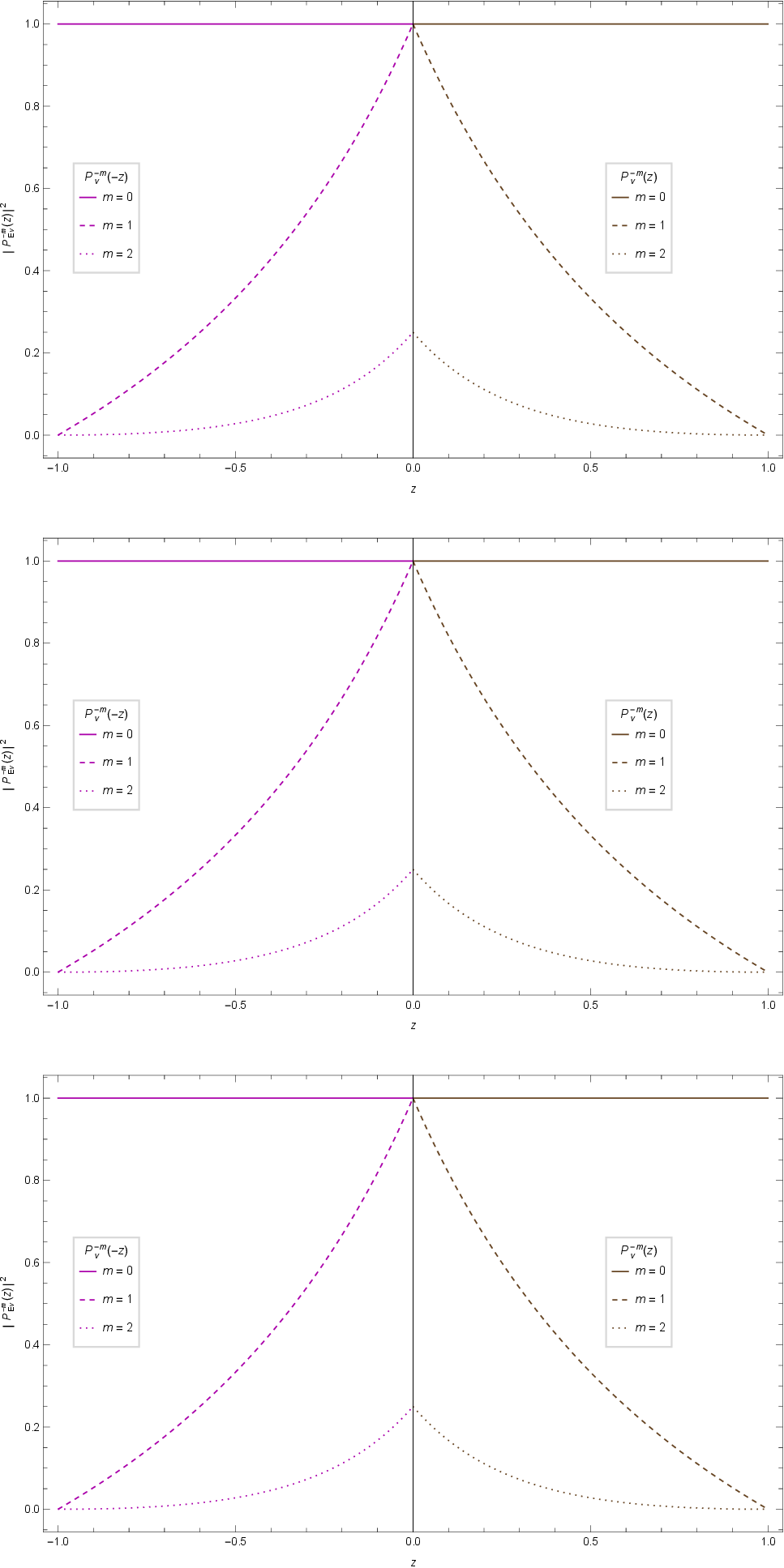}
\caption{\textbf{Top} panel: the first three squared electromagnetic quasibound state angular wave eigenfunctions for $a=0.01$. \textbf{Middle} panel: the first three squared electromagnetic quasibound state angular wave eigenfunctions for $a=0.25$. \textbf{Bottom} panel: the first three electromagnetic squared quasibound state angular wave eigenfunctions for $a=0.49$. We focus on the fundamental mode $n=0$.}
\label{fig:Fig6_4DEGBBH}
\end{figure}

\begin{figure}[h]
\includegraphics[scale=0.6]{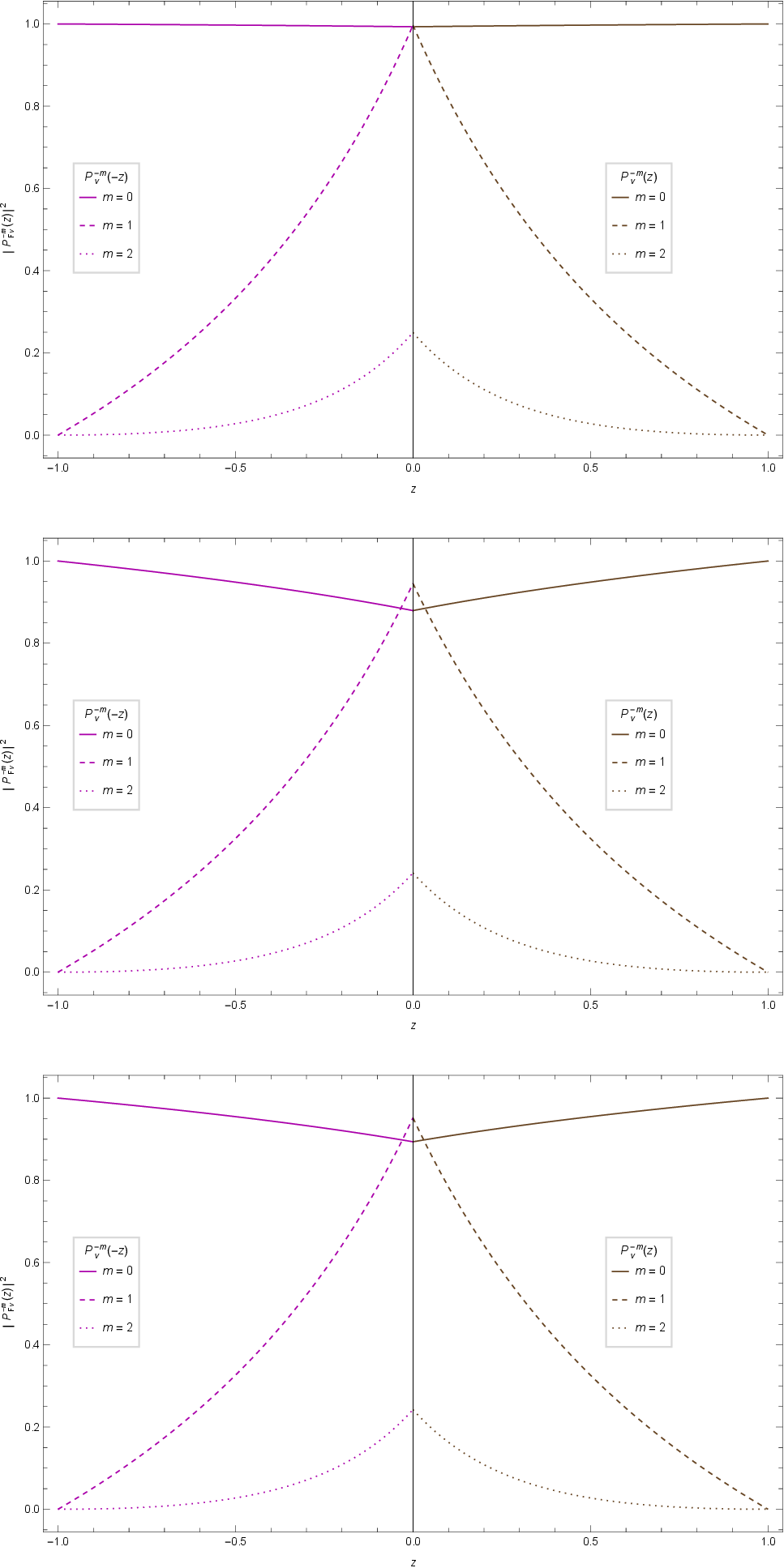}
\caption{\textbf{Top} panel: the first three squared Dirac quasibound state angular wave eigenfunctions for $a=0.01$. \textbf{Middle} panel: the first three squared Dirac quasibound state angular wave eigenfunctions for $a=0.25$. \textbf{Bottom} panel: the first three Dirac squared quasibound state angular wave eigenfunctions for $a=0.49$. We focus on the fundamental mode $n=0$.}
\label{fig:Fig7_4DEGBBH}
\end{figure}

For the fundamental mode $n=0$, the eigenvalues $q_{s0;\eta}$ must obey the relation $c_{1}=0$, where $c_{1}=q/b\gamma$. Thus, we have $q_{s0;\eta}=0$, which implies that there exist only on solution for $q$, namely, $q_{s0;0}=0$, and hence we can obtain the values for the angular eigenvalues $\lambda_{s0;+}$. They are presented in Table \ref{tab:I_4DEGBBH}. As it was expected, the angular eigenvalues $\lambda_{s0;+}$ are in the set of real numbers $\mathbb{R}$, and hence the eigenvalues for the associated Legendre functions can be written as $\lambda=\nu(\nu+1)$. Therefore, we present the behavior of the quasibound state angular wave eigenfunctions in Figures~\ref{fig:Fig5_4DEGBBH}--\ref{fig:Fig7_4DEGBBH}, for the scalar, electromagnetic, and Dirac fields, respectively, as functions of the new angular coordinate $z=\cos\theta$ for some values of the GB coupling constant $a$. It is worth calling attention to fact that (i) the numerically satisfactory solutions of the associated Legendre equation of general (includng complex) degree $\nu$ are given in terms of the Ferrers functions of the first kind $P_{\nu}^{-m}(-z)$ and $P_{\nu}^{-m}(z)$, in the interval $-1 < z < 1$, and (ii) these solutions are regular at the two boundaries $\theta=0$ ($z=1$) and $\theta=\pi$ ($z=-1$).

%
%
\section{Final Remarks}\label{Conclusions}
In this work, we obtained exact analytical solutions of the master wave equations for the test fields in a 4-dimensional Einstein--Gauss--Bonnet black hole spacetime, where the radial solution is given in terms of the general Heun functions, and the angular solution is given in terms of the associated Legendre (Ferrers) functions (of the first kind).

We imposed two boundary conditions on the radial solution in order to study its asymptotic behaviors, which led to the quasibound state phenomena. Near the exterior event horizon, the radial solution describes ingoing waves, which reachs a maximum value and then crosses into the black hole. On the other hand, far from the black hole at asymptotic infinity, the radial solution tends to zero, that is, the probability of finding any particles in the spatial infinity is null.

The spectrum of quasibound states for the test fields was obtained by using the polynomial condition of the general Heun functions. In fact, that is a new (analytical) approach developed by Vieira, Bezerra, and Kokkotas \cite{AnnPhys.373.28,PhysRevD.104.024035}. It is worth pointing out that these massless resonant frequencies were obtained directly from the general Heun functions, and, to our knowledge, there is no similar result in the literature for the background under consideration. In addition, it is worth emphasizing that all the numerical/graphical computations performed in this work were carried out by using the standard package for the general Heun functions installed in Wolfram Mathematica 12.3.

Finally, we have discussed the stability of the system. All the systems are stables in the fundamental mode, and present an overdamped motion, in the range $0 < a < 1/2$. We hope that our results, which describe an unquestionably phenomenon associated with purely quantum effects in gravity, may be used to fit some astrophysical data in the near future, as for example, the ones related to the observations of some spectrum of thin accretion disks with present and future X-ray facilities \cite{PhysRevD.95.104043}, as well as from the secondary object in GW190814 which is compatible with being a slowly-rotating neutron star in EGB theory of gravity \cite{JCAP.02.033}, and hence shed some light on the physics of black holes and compact objects.

As a future perspective, it is possible to extend our results for the case of an asymptotically de Sitter spacetime, as well as for negative values of the GB coupling constant. In addition, we can obtain a new acoustic curved black hole embedded in the 4DEGB spacetime, which could be a very interesting framework within the analog models of gravity.
%
%
%
%
%
\section*{Data availability}
The data that support the findings of this study are available from the corresponding author upon reasonable request.
%
%
\begin{acknowledgments}

The author was funded by the Alexander von Humboldt-Stiftung/Foundation (Grant No. 1209836). This study was financed in part by the Coordena\c c\~{a}o de Aperfei\c coamento de Pessoal de N\'{i}vel Superior - Brasil (CAPES) - Finance Code 001. This study was financed in part by the Conselho Nacional de Desenvolvimento Cient\'{i}fico e Tecnol\'{o}gico - Brasil (CNPq) - Research Project No. 150410/2022-0. The author would like to thank Professor K. D. Kokkotas and Dr. Kyriakos Destounis for useful discussions on topics covered in this work. It is a great pleasure to thank the Theoretical Astrophysics at T\"{u}bingen (TAT Group) for its wonderful hospitality and technological support (software licenses).

\end{acknowledgments}
%
%

%
%
\end{document}